\documentclass[a4paper,12pt]{article}
\usepackage[utf8]{inputenc}
\usepackage{amsmath}
\usepackage{amsfonts}
\usepackage{amssymb}
\usepackage{amsmath, amssymb, graphics, setspace}
\usepackage{graphicx}

\title{Spherically Symmetric Solution in Bi-metric theory of Gravity }
\author{{$\hbox{Anoop Narayanan P E }^{*}$  and $\hbox{P K Suresh}^{\dagger}$}\\
 \begin{small}School of Physics, University of Hyderabad, Hyderabad - 500046, India.\end{small}\\
 $^*$ \begin{small}anoopn@uohyd.ac.in\end{small}\\
 $^\dagger$\begin{small}pkssp@uohyd.ernet.in\end{small}}
\date{}
\begin{document}
\maketitle
\begin{abstract}
The possibility of spherically symmetric solutions in bi-metric theory of gravity is examined. It is shown that two possible black hole type solutions exists in the model. Spherically symmetric solution of general theory of relativity is recovered in the absence of the second metric. The result is compared with  other bi-metric models as well as general theory of relativity.
\end{abstract}

\section{Introduction}
It is proposed that the mathematical formalism of General Theory of Relativity(GTR) can be improved by introducing an additional metric  corresponding to a flat space time \cite{{Rosen1},{Papapetrou},{Rosen}}. The formalism satisfies general covariance  and equivalence principles and keeps all the physical predictions of GTR the same. One of the  advantages in this scenario is that several quantities like gravitational energy and momentum density that was not having tensor form acquires tensor form \cite{Rosen1}. This bi-metric model does not involve coupling between the two metrics. For convenience we call it as BM-I.
 \par
 Attempts have been made to couple the two metrics through a field. The field may be a vector field \cite{Clay3} or a scalar field \cite{Clayton2}. A simplified model with a real scalar field is proposed  with out violating positivity conditions of energy density and pressure and with a zero cosmological constant\cite{Clayton2}. We refer to it as BM-II. Ever-since its introduction various attempts have been made to solve numerous cosmological puzzles \cite{{puzzle},{Early},{Gopakumar}} but very little had been done to explore spherically symmetric (SS) solutions in this model. SS solution has by way of its elegant simplicity found place among the problems of fundamental importance. Attempts has been made to find out black hole type solution in BM-I scenario \cite{{Rosen1},{Mark},{Karade}}. In BM-I it is shown that SS solution does exist but it does not involve any black hole type solution\cite{Rosen}. 
 \par
As mentioned earlier extensive study is done relating to the application of BM-II to cosmology, but little had been done to work out SS solutions in this model. Therefore the aim of the present study is to investigate the possibility of SS solution in BM-II. We show that SS solutions exists in this model with black hole type solution. Further it is shown that the gravitational radius acquires a bi-metric feature and from which single metric form of GTR can be recovered.\\
 In the second section we discuss BM-II formalism briefly. In the third section we calculate the vacuum field solution for SS distribution of matter. Discussion and conclusion are presented in the last section.
  
 \section{Bi-metric theory of gravity}
 Various models of bi-metric theories has been proposed\cite{{Rosen1},{Papapetrou},{Rosen},{other1},{other}}. In the present study we work with BM-II as given by \cite{{Clayton2},{Clayton}},
 \begin{equation}\label{eq:bim}
\hat {g}_{\mu\nu}=A[\phi]g_{\mu\nu}+B[\phi]\partial_\mu\phi\partial_\nu\phi.
\end{equation}

Where $ g_{\mu\nu} $ is the gravitational metric, $\hat {g}_{\mu\nu}$ is  the matter metric and $\phi$ is called as the biscalar field. The field $\phi$ is assumed to be  minimally coupled  to the gravitational field described by $g_{\mu\nu}$ which is governed by Klein- Gordon equation and not necessarily same as other scalar field appearing in  matter model. The inverse metrics $\hat {g}^{\mu\nu} $and $g^{\mu\nu}$ satisfy the following relations

\begin{equation}
\hat {g}^{\mu \alpha}\hat g_{\nu \alpha}=\delta^{\mu}_{\nu}\,\,\, \text{and}\,\,\, {g}^{\mu \alpha}g_{\nu \alpha}=\delta^{\mu}_{\nu}.
\end{equation}
One recovers GTR when $A[\phi]=1$ and $ B[\phi]=0$. In the present study we take  $A[\phi]=1$ and $ B[\phi]=B= \text{constant}$\cite{Clayton}. Thus (\ref{eq:bim}) becomes
 \begin{equation}\label{bm}
\hat {g}_{\mu\nu}=g_{\mu\nu}+B\partial_\mu\phi\partial_\nu\phi.
\end{equation}
  Where $B\simeq \frac{1}{32\Pi}l_p^2$ and $l_p$ is the Planck length\cite{Clayton}. The key motivation behind the choice of $B$ is that it gives the correct order of magnitude of CMB spectrum\cite{Moffat}. This model is applied to Friedmann Lema$\hat{i}$tre Robertson Walker (FLRW) cosmology \cite{Early}. In the current study the biscalar field is taken as homogenous scalar field.

\section{Vacuum field solution}
The goal of the present section is to obtain the vacuum field solution for a spherically symmetric homogenous distribution of matter.
Consider a centrally symmetric gravitational field produced by a centrally symmetric distribution of matter. This means that the line element $ds^2$ should have same value for all the points located at equal distance from the centre. Thus  the most general form of expression for the line element  can be written as,

\begin{equation}\label{ds2}
ds^2={\cal{P}}(r,t)dt^2+{\cal{Q}}(r,t)dr^2+{\cal{R}}(r,t)r^2(\sin^2 \theta d\phi^2+d\theta^2)+{\cal{S}}(r,t)drdt.
\end{equation}
One can perform the following a co-ordinate transformation, 
\begin{equation}
r' = r '(r, t)\,\,\,\, \text{and}\,\,\,\, t' = t '(r, t),
\end{equation}
which does not destroy the central symmetry of the system.
This  permit us to have a choice of the values for any two quantities among $\cal{P},\cal{Q},\cal{R},\cal{S}$ and $\cal{T}$. We may chose $\cal{R}$=-1 and $\cal{S}$=0. Thus (\ref{ds2}) gives,\footnote{Here onwards prime is omitted for convenience}

\begin{equation}\label{ds22}
ds^2={\cal{P}}(r,t)dt^2+{\cal{Q}}(r,t)dr^2-r^2(\sin^2\theta d\phi^2+d\theta^2)
\end{equation}
It is convenient to write $\cal P$ and $\cal Q$ in the exponential form as $c^2e^{\nu}$ and $-e^{\lambda}$ respectively. Thus (\ref{ds22}) yields
\begin{equation}\label{line}
ds^2=e^{\nu}c^2dt^2-e^{\lambda}dr^2-r^2(d\theta ^2+\sin^2\theta d\phi^2).
\end{equation}
In the bi-metric scenario with (\ref{bm}) the line element (\ref{line}) takes the following form.
\begin{equation}
ds^2=(e^{\nu}+B\dot{\phi}^2)c^2dt^2-e^{\lambda}dr^2-r^2(d\theta ^2+\sin^2\theta d\phi^2).
\end{equation}
Thus, the non zero components of the metric $\hat{g}_{\mu \nu}$ and its inverse are respectively given by:\\
$~~~~~\hat{g}_{00}=e^{\nu}+B\dot{\phi}^2$~~~~~$\hat{g}_{11}=-e^{\lambda}$~~~~~$\hat{g}_{22}=-r^2$~~~~~$\hat{g}_{33}=-r^2 \text{sin}^2\theta$,\\

$\hat{g}^{00}=(e^{\nu}+B\dot{\phi}^2)^{-1}$~~~~~$\hat{g}^{11}=-e^{-\lambda}$~~~~~$\hat{g}^{22}=-r^{-2} $~~~~~$\hat{g}^{33}=-r^{-2} \text{sin}^{-2}\theta$.\\
The nonvanishing Christoffel symbols are computed and are given below\\
\begin{doublespace}
\noindent\(\begin{array}{llllll}
\text{$\Gamma^{0}_{1 1} $}= & \frac{e^{\lambda } \dot{\lambda}} {2 \left(e^{\nu }+B \dot{\phi }^2\right)} &
 \text{$\Gamma^{0}_{01} $}= & \frac{e^{\nu } \nu '}{2 \left(e^{\nu }+B \dot{\phi}^2\right)} &
 \text{$\Gamma^{0}_{0 0} $}= & \frac{2 B\dot{ \phi } \ddot{\phi }+e^{\nu } \dot{\nu}}{2 e^{\nu }+2 B \dot{\phi} ^2}\\ 
 
 \text{$\Gamma^{1}_{1 1} $}= & \frac{1}{2} \lambda' & \text{$\Gamma^{1}_{2 2} $}= & -r e^{-\lambda } &\text{$\Gamma ^{1}_{3 3}$}= & -e^{-\lambda } r \text{sin}^2\theta  \\

 \text{$\Gamma^{1}_{0 1} $}= & \frac{\dot{ \lambda}}{2}  &
 \text{$\Gamma^{1}_{0 0} $}= & \frac{\nu '}{2} e^{-\lambda +\nu } &
 \text{$\Gamma ^{2}_{2 1}$}= & \frac{1}{r} \\
 \text{$\Gamma^{2}_{3 3} $}= & -\text{cos}\theta \, \text{sin}\theta  &
 \text{$\Gamma^{3}_{3 1} $}= & \frac{1}{r} &
 \text{$\Gamma^{3}_{3 2} $}= & \text{cot}\theta  .\\
 
\end{array}\)
\end{doublespace}
Here dot denotes derivative with respect to $t$ and prime denotes derivative with respect to $ r$.
This help us to calculate the components of Einstein tensor and the field equations are given as follows,
\begin{eqnarray}\label{bim1}
G_{0}^{0}&= &\frac{e^{-\lambda}(e^{\lambda}-1+r\lambda ')}{r^2}=\frac{8\pi G }{c^4}T_{0}^{0}\\
G_{0}^{1}&=&\frac{-e^{-\lambda}\dot{\lambda}}{r}=\frac{8\pi G}{c^4}T_{0}^{1}\\
G_{1}^{1}&= & \frac{-B \left(-1+e^{\lambda }\right) \dot{\phi}^2+e^{\nu } \left(1-e^{\lambda }+r \nu '\right)}{r^2
\left(e^{\nu }+B \dot{\phi} ^2\right)}\nonumber \\
&=&\frac{8\pi G}{c^4}T_{0}^{0}\\
G_{2}^{2}&=&G_{3}^{3}\nonumber \\
&=&\frac{r^2}{4(e^{\nu}+B\dot{\phi}^2)} re^{-\lambda}\Big (-2Be^{\lambda}r\dot{\phi}\ddot{\phi}\dot{\lambda}+2B^2\dot{\phi}^4\lambda'+e^{\nu}(e^{\lambda}r\dot{\lambda}^2 \nonumber\\
& & -e^{\lambda}r\dot{\lambda}\dot{\nu}+2e^{\lambda}r\ddot{\lambda}+e^{\nu}(-2\nu'-r\nu'^2+\lambda'(2+r\nu')-2r\nu''))\nonumber\\
& &+B\dot{\phi}^2\Big (re^{\lambda}\dot{\lambda}^2+2e^{\lambda }r\ddot{\lambda}+e^{\nu}(\lambda'(4+r\nu')-2(\nu'+r\nu'^2+r\nu''))\Big)\Big)\nonumber\\
&=&\frac{8\pi G}{c^4}T_{3}^{3}=\frac{8\pi G}{c^4}T_{2}^{2} .
\end{eqnarray}
Here $T_{\mu}^{\nu}$ represents the energy momentum tensor. Thus vacuum field solution for SS distribution of matter can be written using (9)-(12) as follows
\begin{eqnarray}\label{da}
\dot{\lambda}&=&0\\
\frac{e^{-\lambda}\lambda'}{r}-\frac{e^{-\lambda}}{r^2}+\frac{1}{r^2}&=&0\\
\frac{e^{\nu-\lambda}\nu'}{r(e^{\nu}+B\dot{\phi}^2)}+\frac{e^{-\lambda}}{r^2}-\frac{1}{r^2}&=&0.
\end{eqnarray}
The equation (\ref{da}) implies that $\lambda$ has only $r$ dependance. Adding the equations (14) and (15) give the relation between $\lambda  $ and $\nu  $ as\\

\begin{equation}
\partial_{r}\big(\lambda+\ln(e^{\nu}+B\dot{\phi}^2)\big)=0,
\end{equation}
where $\partial_{r}$ denotes partial derivative with respect to $r$. This  allow us to establish the relation between $\nu$ and $\lambda$ as\\
\begin{equation}
e^{\lambda}=\frac{1}{e^{\nu}+B\dot{\phi}^2}.
\end{equation}
Using equation (15) we get
\begin{equation}
e^{-\lambda}=e^{\nu}+B\dot{\phi}^2=1+\frac{F}{r},
\end{equation}
where $F$ is a constant.
We motivate the choice of $F$ based on the fact that in  the absence of the second metric  we should recover the Newtonian limit and also $\hat{g}_{00}$ should reduce to Galilean  in the limit when $r\rightarrow \infty$. $F$ can accommodate an additional term $\pm B \dot{\phi}^2$ without violating equations (13-15). Therefore two different solutions  emerge in this scenario. This can be substituted in (8) and we obtain SS solution for the bi-metric model as,

\begin{equation}
ds^2=(1-\frac{2GM}{rc^2}\pm \frac{B\dot{\phi}^2}{r})c^2dt^2-\frac{dr^2}{(1-\frac{2GM}{rc^2}\pm \frac{B\dot{\phi}^2}{r})}-r^2(d\theta^2 + \sin ^2\theta d\phi ^2).
\end{equation}

Thus the gravitational radius of SS distribution in BM-II formalism can be written as,
\begin{equation}
r_{BM}=\frac{2GM }{c^2}\mp B\dot{\phi}^2.
\end{equation}
This shows that the black hole type solution exist in the bi-metric scenario. In the absence of $B$ (the second metric) the results matches with corresponding solution in GTR. 
\par

Now we compare the result (20) with the models of \cite{Rosen}, \cite{Mark} and \cite{Karade}. The SS solutions in all the above models are tabulated in Table \ref{t1}.

\begin{table}[ht]
\caption{\label{t1}Comparison of spherically symmetric solutions in various bi-metric models and GTR}
\centering
\begin{tabular}{|l|l|l|}
\hline
Theories&Spherically Symmetric Line Element &Gravitational Radius \\
\hline
GTR&$ds^2=(1-\frac{2GM}{rc^2})c^2dt^2$&\\
 &\,\,\,\,\,\,\,\,\,\,\,\,\,\ $-\frac{dr^2}{(1-\frac{2GM}{rc^2})}-r^2(d\theta^2 + \sin ^2\theta d\phi ^2)$&$\displaystyle{r=\frac{2GM}{c^2}}$\\
\hline
  &$ds^2=e^{-2m/r}c^2dt^2$&\\
  &\,\,\,\,\,\,\,\,\,\,\,\,\,\ $-e^{2m/r}(dr^2+d\theta^2+\sin^2\theta d\phi^2)$\cite{Rosen}&\\
BM-I&$ds^2=-dr^2-r^2(d\theta ^2+\sin^2\theta d\phi^2)+$&No blackhole\\
&\,\,\,\,\,\,\,\,\,\,\,\,\,\,\,\,$e^{-\alpha/r}c^2dt^2$\cite{Karade}&type solution \\
&$ds^2=-e^{\alpha /r}(dr^2+r^2(d\theta^2+\sin^2\theta d\phi^2))$&\\
&\,\,\,\,\,\,\,\,\,\,\,\,\,\,\,\,$+e^{\beta/r}c^2dt^2$\cite{Karade}&\\
\hline
BM-II&$ds^2=(1-\frac{2GM}{rc^2}\pm \frac{B\dot{\phi}^2}{r})c^2dt^2$&\\
  &$\,\,\,\,\,\,\,\,\,\,\,\,\,\ -\frac{dr^2}{(1-\frac{2GM}{rc^2}\pm \frac{B\dot{\phi}^2}{r})}-r^2(d\theta^2 + \sin ^2\theta d\phi ^2)$&$r_{BM}=\frac{2GM }{c^2}\mp B\dot{\phi}^2$\\
\hline
\end{tabular}
\end{table}
All the bi-metric models exhibit the feature of SS vacuum field solution and they agree to GTR result in the absence of the second metric. In the case of BM-II the black hole type solution exists in contrast to other bi-metric models. Gravitational radius of the model under consideration reduces to Schwarzschild solution in the absence of biscalar field $\phi$.

\section{Conclusion}
We examined vacuum field solution for SS distribution of matter in BM-II. It is shown that two SS solutions are possible. The metric components in BM-II  gives reduces to Galilean  form in the limit $r\rightarrow\infty$. Single metric form of GTR can be recovered in the absence of $B$. The analysis of the present result shows that black hole type solution exists in BM-II in contrast to BM -I. The additional correction ie, $\pm B \dot{\phi}^2$ to Schwarzschild radius is entirely due to the presence of second metric. The result may be used to study  the solar system scale phenomenon and may verify the feasibility of bi-metric scenario  which are beyond the scope of the present work.

\section*{Acknowledgements}
One of the author ANPE would like to thank for providing financial support through BBL, University of Hyderabad.


\begin{thebibliography}{00}
\bibitem {Rosen1} Rosen N 1940 $\it {Phys. Rev.}$  {$\bf 57$} 147
\bibitem{Papapetrou} Papapetrou A 1948 $\it{Proc. Roy. Irish\, Acad}$ $\bf{52}$ 11
\bibitem{Rosen} Rosen N 1973 $\it {Gen.Rel.Grav}$ {$\bf 4$} 435
\bibitem{Clay3} Clayton  M A and  Moffat J W 1999 $\it{Phys. Lett. B}$ {$\bf 460$} 263
\bibitem{Clayton2}  Clayton M A and Moffat J W  2000 $\it{Phys. Lett. B}$ {$\bf 477$} 269
\bibitem{puzzle} Albrecht A and Magueijo J 1999  $\it{Phys. Rev. D}$ {$\bf 59$} 043516
\bibitem{Early} Thom$\acute{a}$s J G arXiv:1312.3548v1 [gr-qc]
\bibitem{Gopakumar}  Gopakumar P and Vijayagovindan G V  2006 $\it{Int.J.Mod.Phys.}$ D {$\bf  15$} 321
\bibitem {Mark}  Israelit M 1981 $\it{Gen.Rel. Grav.}$ {$\bf13$} 681
\bibitem{Karade} Karade T M 1980 $\it{Indian\,J.pure\, appl.\, Math}$ {$\bf 11$} 1202
\bibitem{other1} Lightman and Lee 1973 $\it{Phys. Rev. D}$ {$\bf{8}$} 3293
\bibitem{other} Clifton et.al arXiv : 1106.2476v3 [astro-ph]
\bibitem {Clayton}  Clayton M A and  Moffat J W  2001 $\it {Phys. Lett. B}$ {$\bf 506$}   177
\bibitem{Moffat} Moffat J W  arXiv : 1306.5470v1 [gr-qc]
\end{thebibliography}
\end{document}